# Title: pH-conditioning of recognition layers enables single-molecule affinity detections at $10^{-20}$ molar

**Authors:** Eleonora Macchia[1,2,3]†, Cinzia Di Franco[4,2]†, Cecilia Scandurra[5,2], Lucia Sarcina[5,2], Matteo Piscitelli[6], Michele Catacchio[1,2], Mariapia Caputo[1,2], Paolo Bollella[5,2], Gaetano Scamarcio[6,4,2]*, and Luisa Torsi[5,2]*

**Affiliations:** [1]Dipartimento di Farmacia-Scienze del Farmaco, Università degli Studi di Bari Aldo Moro, 70125 Bari, Italy; [2]Centre for Colloid and Surface Science, Dipartimento di Chimica, Università degli Studi di Bari Aldo Moro, 70125 Bari, Italy; [3] Faculty of Science and Engineering, Åbo Akademi University, Turku, Finland; [4]CNR IFN, 70126 Bari, Italy, [5]Dipartimento di Chimica, Università degli Studi di Bari Aldo Moro, 70125 Bari, Italy; [6]Dipartimento Interateneo di Fisica, Università degli Studi di Bari Aldo Moro, 70125 Bari.

†Equally contributing first authors
*Corresponding authors Email: luisa.torsi@uniba.it, gaetano.scamarcio@uniba.it

**Abstract:** While nucleic-acids can be readily amplified for single-marker detection, a comparable method for proteins assay is currently unavailable. Proteins potentiometric detections at $10^{-20}$ molar have been demonstrated, but the mechanism remains elusive. Here, we unveil how pH-conditioning within the trillions of recognition elements densely packed on a millimeter-large surface, enables single protein or DNA selective detections in 0.1 mL of a biofluid. Plasmonic, electronic and surface probing techniques demonstrate that a conformational change, elicited by a single-affinity binding, alters the secondary and tertiary structure of the recognition elements. A phenomenological mechanism foresees that the pH-conditioning initiates a hydrophobization process leading to the formation of a partially aggregated and metastable state that facilitates the amplification spreading. Impact on protein aggregates control and biomarker-based diagnostics, is envisaged.

**One-Sentence Summary:** A mechanism shows how pH shifts enable to detect a conformational change spreading over $10^{12}$ capturing antibodies or probes

Early-stage diagnosis, essential for improving therapeutic outcomes, often relies on the highly sensitive detection of biomarkers in complex peripheral body fluids such as serum or urine (*1*, *2*). Polymerase Chain Reaction (PCR) (*3*), capable of amplifying a single copy of a nucleic acid marker in 0.1 mL ($10^{-20}$ mole/L, M), serves as a cornerstone in molecular diagnostics. It enables rapid, reliable, and highly sensitive detections. Additionally, PCR plays a key role in genome profiling (*4*), supporting the sequencing of DNA and RNA targets, including novel transcripts. While the significance of protein markers assays, alone or as complement to genetic markers, for enhanced diagnostic accuracy (*5*, *6*) is acknowledged, there is currently no amplification technology available for protein assays. This limitation persists despite the potential for single-molecule protein sequencing (*7–10*) that, while technically achievable, remains impractical for routine rapid tests. Similarly, digital beads-based assays (*11*), typically operating in the $10^{-15}$ M range but capable of reaching $10^{-19}$ M (*5*), are bench-top systems with time-to-results of several hours. A potentiometric platform qualitatively detects proteins and nucleic acids at $10^{-20}$ M within an hour (*12*, *13*). Tests in clinical settings demonstrate a diagnostic sensitivity exceeding 96% in pancreatic cancer patients (*14*, *15*), which opens new possibilities for point-of-care screening for early diagnosis. Nonetheless, a sensing mechanism is required to improve consistency even further.

From a more fundamental perspective, the detection of single proteins typically relies on their spatial-temporal confinement within a near-field volume ranging from $10^{-18}$ to $10^{-15}$ L (*16*), such as within a microwell (*11*) or a nanopore (*9*). This approach allows for an increase in the concentration of the target protein while minimizing interference from other, more abundant, species present in the complex biofluid. Techniques for detecting this near-field response include, scanning probe microscopy (*17*), sensors based on nanowire field-effect transistors (FETs) (*18*), and single-molecule plasmonic methods (*19*). However, a significant challenge remains: the concentration of the target protein typically falls within the $10^{-9}$ M to $10^{-6}$ M range. Hence, they achieve single-molecule resolution but limit-of-detection, LOD (*20*) remains many orders of magnitudes higher than $10^{-20}$ M.

Here, we study single-molecule proteins and DNA selective detections with an ensemble of plasmonic, electronic and surface probing techniques that reveal the critical role of a pH conditioning within a millimeter large biolayer detecting surface. This is populated by trillions of highly packed capturing antibodies or protein-probe complexes recognition elements. The pH-shift enabling role in detections at $10^{-20}$ M, is proven through studies involving Surface Plasmon Resonance (SPR) (*21*), electrolyte-gated organic FET (EGOFET) (*22*) and Kelvin Probe Force Microscopy (KPFM) (*23*). A phenomenological mechanism is proposed also considering the increase of the Static Contact Angle (SCA) by almost 20 (°). Further support to the hydrophobization, comes from nanomechanical investigations, which reveal that the surface becomes over 60 % more adhesive. An amplification mechanism stemming from a pH-induced partial unfolding of the recognition elements, ultimately exposing some of their hydrophobic regions, is proposed. Such a conditioning, initiating an aggregation process (*24*) through short-range hydrophobic interactions (*25*), is likely to place the recognition elements biolayer in a metastable state. A Polarization Modulation Infrared Reflection-Absorption Spectroscopy (PM-IRRAS) study focused on the amide spectral regions, demonstrates that single-molecule affinity binding alone already induces a change in the secondary structure, while pH-conditioning enables the rearrangement to reach the tertiary structure. Both structural modifications start from a single affinity couple conformational electrostatic change and entails a spreading over at least hundreds of millions of the packed recognition elements, but the effect is more pronounced with pH-conditioning, enabling reliable detections at $10^{-20}$ M.

**SPR single/few-molecules affinity sensing**

SPR serves as a highly sensitive probe to inspect ultra-thin layers deposited on a metalized slide (**fig. S1a**). It typically assays at LODs in the $10^{-9}$ M concentration range (*26, 27*), reaching $10^{-18}$ M by exploiting plasmon-enhanced effects (*28–30*). SPR reflectance intensity goes through a dip at the $\Delta\theta_{SPR}$ shift of the laser incident angle, $\theta$, corresponding to the energy-wavevector conservation between the exciting photons and the surface plasmon polaritons (*21*).

The generality of the present study is based on the evaluation of three different biosystems, schematically shown in **figs. S1b-S1d**, namely: *(i)* an anti-HIV-1-p24 capturing layer designed to detect the p24 antigen of the HIV-1 virus capsid, with the non-binding C-reactive protein (CRP) serving as interferer or as target in negative control experiments, *(ii)* an anti-Immunoglobulin G (anti-IgG) layer for detecting IgG, with IgM as non-binding species, and *(iii)* a biolayer of protein-probe complexes comprising a neutravidin (NA) coating coupled to biotinylated complementary KRAS strands, designated as b-KRAS. The whole probe-layer is addressed as NA-b-KRAS. The KRAS mutated gene, used as a marker for pancreatic cancer precursors (*2*), is the target, while TP54, another mutated gene, serves as non-affinity target.

SPR traces that measure transient $\Delta\theta_{SPR}$ shifts *in-situ* and *operando*, are referred to as sensograms. Typical sensograms of the anti-HIV-1-p24, anti-IgG or NA-b-KRAS biolayers depositions on a millimeter large slide, are shown in **figs. S2-S4**. The physisorption occurs in a 4-(2-hydroxyethyl)-1-piperazineethanesulphonic acid (HEPES) buffer at pH 7.4 and ionic strength, $i_s$, of 150 mM (HEPES@pH7.4), mimicking physiological conditions. The process leads to stably adherent monolayers (*31*), which are 5 - 9 nm thick (**tab. S1**) and consist of $10^{11}$-$10^{12}$/cm$^2$ (**tab. S2**) highly packed (*32*) recognition elements.

The SPR signal measured on a as deposited anti-IgG layer exposed to IgG proteins in the of $10^{-7}$ M range (**fig. S5**), is modelled using the refractive index increment factor (*33*). Since a density of IgGs comparable to the anti-IgGs in the capturing layer is attached, we refer to this as the "double-layer" regime. Here, no significant signal is detected below $10^{-9}$ M concentrations. The situation changes drastically when the pristine biolayers are conditioned through exposure to a buffer at a pH of either 6 (HEPES@pH6) or 8 (HEPES@pH8), both with an $i_s$ of 150 mM. Subsequently, to retain only the irreversible changes, the pH is returned to 7.4 by washing in HEPES@pH7.4 (**fig. S6**).

In **Fig. 1** the sensograms on non-conditioned (black curves) and pH-conditioned (red-curves: pH 6, blue-curves: pH 8) capturing or protein-probe biolayers, are shown for the three biosystems. The $1 \cdot 10^{-20}$ M (1 ± 1 target in 0.1 mL) "single-molecule" and the $1 \cdot 10^{-19}$ M (10 ± 3 targets in 0.1 mL) "few-molecules" sensing regimes are investigated, along with the non-binding interferer at $1 \cdot 10^{-15}$ M ($10^5$ molecules in 0.1 mL). The sensing protocol starts with the measure of the baseline in the HEPES@pH7.4 buffer or in serum. Then, these are added with the non-binding interferer and injected in the SPR cell. Afterwards, two subsequent assays of single and of few targets in buffer/serum solutions, are performed. Specifically, **Fig. 1A** shows the sensograms for the HIV-1-p24 protein in HEPES@pH7.4, while **Fig. 1B** shows the same assay carried out in diluted human serum. **Fig. 1C** and **Fig. 1D** pertain to similar experiments conducted on anti-IgG (here multiple conditioning on the same sample is performed) and on NA-b-KRAS, respectively.

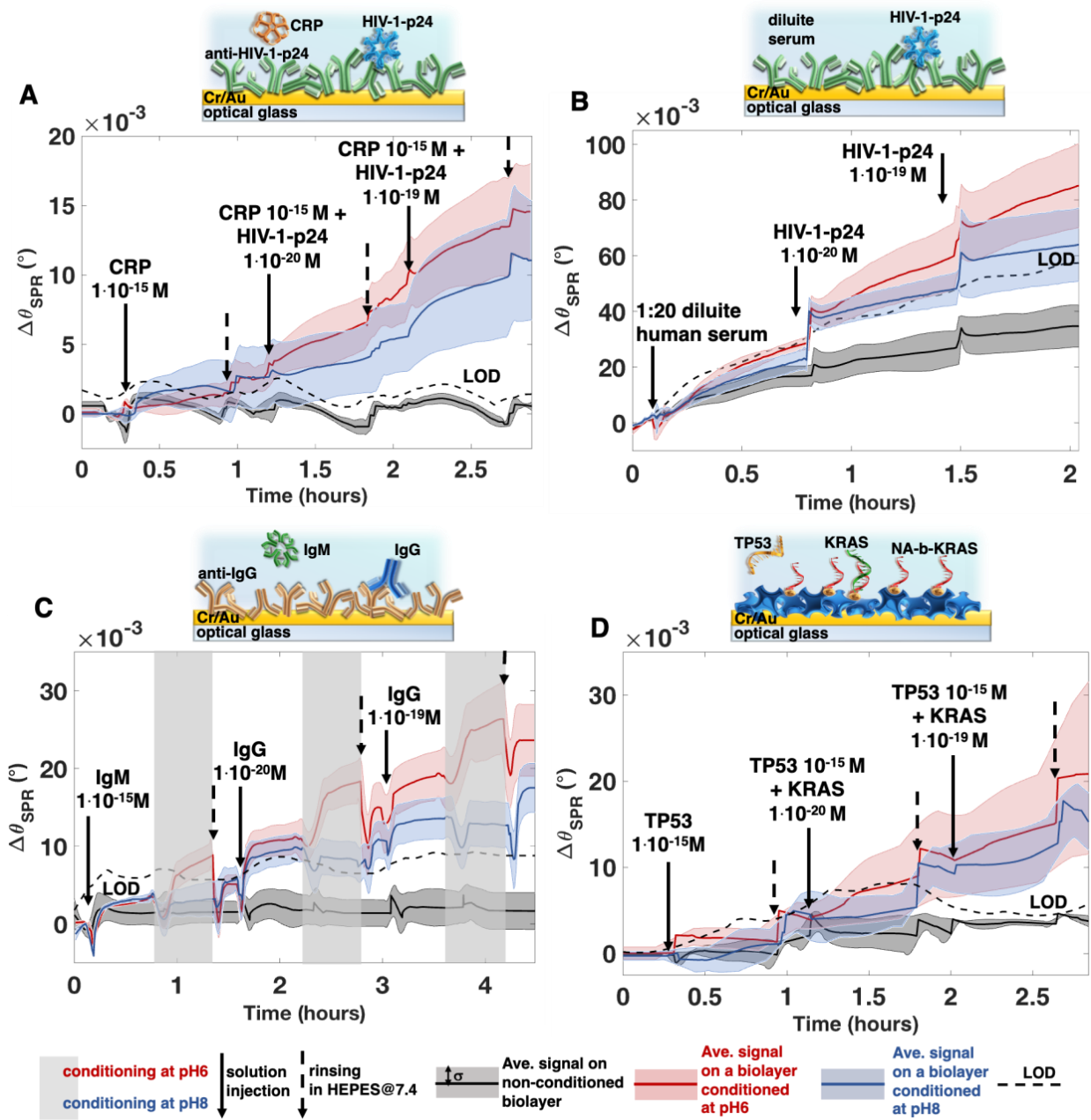

**Fig. 1. SPR single/few-molecules detections of two antigens and a DNA strand in a physiological buffer and in diluted human serum -** Average sensing curves of non-conditioned (black lines) and pH-conditioned biolayers (red lines: pH 6, blue lines: pH 8). The colored shadings represent one standard deviation, while the dashed black lines are the LOD levels, calculated as the average values of each black curves (taken as the measure noise) plus three times the standard deviation. All the solutions are in an HEPES@pH7.4 buffer ($i_s$ = 150 mM) or in diluted human serum. The solid arrows indicate the injection of 1 mL of a given solution, while the rinsing steps are indicated by dashed arrows. (**A**) Anti-HIV-1-p24 capturing layer exposed to the non-binding CRP ($1·10^{-15}$ M) and to HIV-1-p24 ($1·10^{-20}$ M, $1·10^{-19}$ M, 6 replicates, raw data in **figs. S7-S9**). (**B**) Same as in (**A**) but in 1:20 diluted pooled human serum (4 replicates, raw data in **figs. S10-S12**). (**C**) Anti-IgG capturing layer exposed to the non-binding IgM ($1·10^{-15}$M) and to IgG ($1·10^{-20}$M and $1·10^{-19}$M); vertical shadowing indicates pH conditioning carried out before each sensing step (4 or 6 replicates, raw data in **figs. S13-S15**). (**D**) NA-b-KRAS probes exposed to the

non-binding TP53 ($1\cdot10^{-15}$M) and to KRAS ($1\cdot10^{-20}$M and $1\cdot10^{-19}$M, 4 replicates, raw data in **figs. S16-S18**).

Surprisingly, a signal exceeding the LOD is observed when a single-affinity ligand binds to one recognition element among the trillions present on the large sensing surface. This occurs on the layers conditioned at pH 6, while the conditioning at pH 8 provides a reliable response in the few-molecules regime where higher signals are recorded. By contrast, on the non-conditioned layers, no appreciable signal is recorded until a target concentration of $10^{-7}$ M is injected. Importantly, all the sensograms of **Fig. 1** exhibit a response at $10^{-7}$ M (**figs. S7-S18**), thereby providing that the pH-conditioning does not impair the recognition elements affinity binding capability in the double-layer regime. A systematic comparison between the single/few-molecules and the double-layer regimes, evidencing the different features, is provided in **fig. S19**. An experimental design study further demonstrates that exposure of the biolayers to a sole change in ionic strength does not activate any single/few-molecules regime (**figs. S20-S22**, **tab. S3**).

In addition to single-molecule detections, the SPR responses also exhibit very high selectivity. This is evident from the responses falling below the LOD when the interferents are assayed at concentrations, at least, $10^{-4}$-fold higher than the targets. A response lower than the LOD level is also recorded when human serum, without the addition of non-endogenous targets, is assayed. This occurs despite the presence in serum of proteins such as albumin, globulins, or fibrinogen at concentrations in the mM range, which remain remarkably high even when the actual assayed fluid is diluted 1:20.

The refractive index accounts additively for a protein amino acid content but is also a measure of a change in polarization, **P**, (*33*) associated with structural conformational changes (*34*), or surface charges ($\sigma_S$) (*35*) present on a biolayer. Here we prove that the discovered SPR single/few-molecules regime associated with conformational rearrangements in the capturing/probe layer, also shifts the plasmonic resonance.

**pH conditioning enables single/few-molecules detection *via* a surface potential shift**

Conformational changes associated with affinity bindings can also be detected with field-effect transistors (FETs). The gate/biolayer work function changes result in threshold voltage ($V_T$) and source-drain current ($I_D$) shifts (*36*, *37*). Lately, single-molecules have been potentiometrically detected at $10^{-20}$ M with a millimeter-wide EGOFET (*12*), but the role of the pH enabling factor and the associated sensing mechanism was not disclosed.

In EGOFETs (**fig. S23a**) $V_T$ (*22*), taken as the flat band potential (*38*), is directly correlated to the gate surface potential, $\Phi_S$, and $\sigma_S$ (*12*). The $I_D$ current drifting in the semiconductor channel contacted through the source (S) and drain (D) electrodes, is the measured signal. The FET-channel is capacitively coupled to the gate electrode (G), functionalized with a biological recognition layer (anti-HIV-1-p24, **fig. S23b** or NA-b-KRAS, **fig. S23c**), *via* charge double-layers. A reference electrode is also shown in **fig. S23d**. In the present study, a low salinity HEPES buffer at pH 7.4 and $i_s$ of 5 mM (HEPES@pH7.4/$i_s$-low) serves as the dielectric medium offering, at a physiological pH, a Debye Length ($\lambda_D$) of 4.3 nm comparable to the recognition layer thickness. This assures adequate unscreening of the gate electrode $\sigma_S$ with sufficiently high induced $I_D$ current.

In **Fig. 2A** and **Fig. 2B** (raw data in **figs. S24-S26**) the $I_D$ and $V_T$ shifts for HIV-1-p24 sensing on differently conditioned anti-HIV-1-p24 capturing layers, are shown at different steps of the single/few-molecules sensing. In **Fig. 2C** and **Fig. 2D** (raw data in **figs. S27-S29**) the same data are presented for the KRAS sensing. All the measurements involve the incubation of the biofunctionalized gate in the HEPES@pH7.4 buffer at physiological salinity ($i_s$ = 150 mM), added with the non-binding species at $1·10^{-15}$ M or with a single ($1·10^{-20}$ M) or few ($1·10^{-19}$ M) affinity ligands. After incubation, the $I_D$ current measurements are performed in HEPES@pH7.4/$i_s$-low, hence the non-conditioned electrodes are never exposed to a pH shift. Remarkably, the data for the non-conditioned electrodes (black hollow squares) all fall below the LOD, clearly indicating that the FET-response is inhibited when no pH-conditioning occurs. By contrast, the gates conditioned at pH 6 (red hollow circles) or at pH 8 (blue hollow circles), display a reliably detectable signal beyond the LOD, already at single-molecule binding. Quantitatively, the $\Delta V_T$ registered at $10^{-19}$ M on the anti-HIV-1-p24 layer conditioned at pH 6, is (28 ± 3) mV. A $\Delta V_T$ of (63 ± 18) mV is measured on the anti-HIV-1-p24 covered gate as compared to the bare gold one. Hence, a relative $\Delta V_T$ increase of (44 ± 14) %, with respect to the biofunctionalized gate, upon few-molecules sensing, is measured.

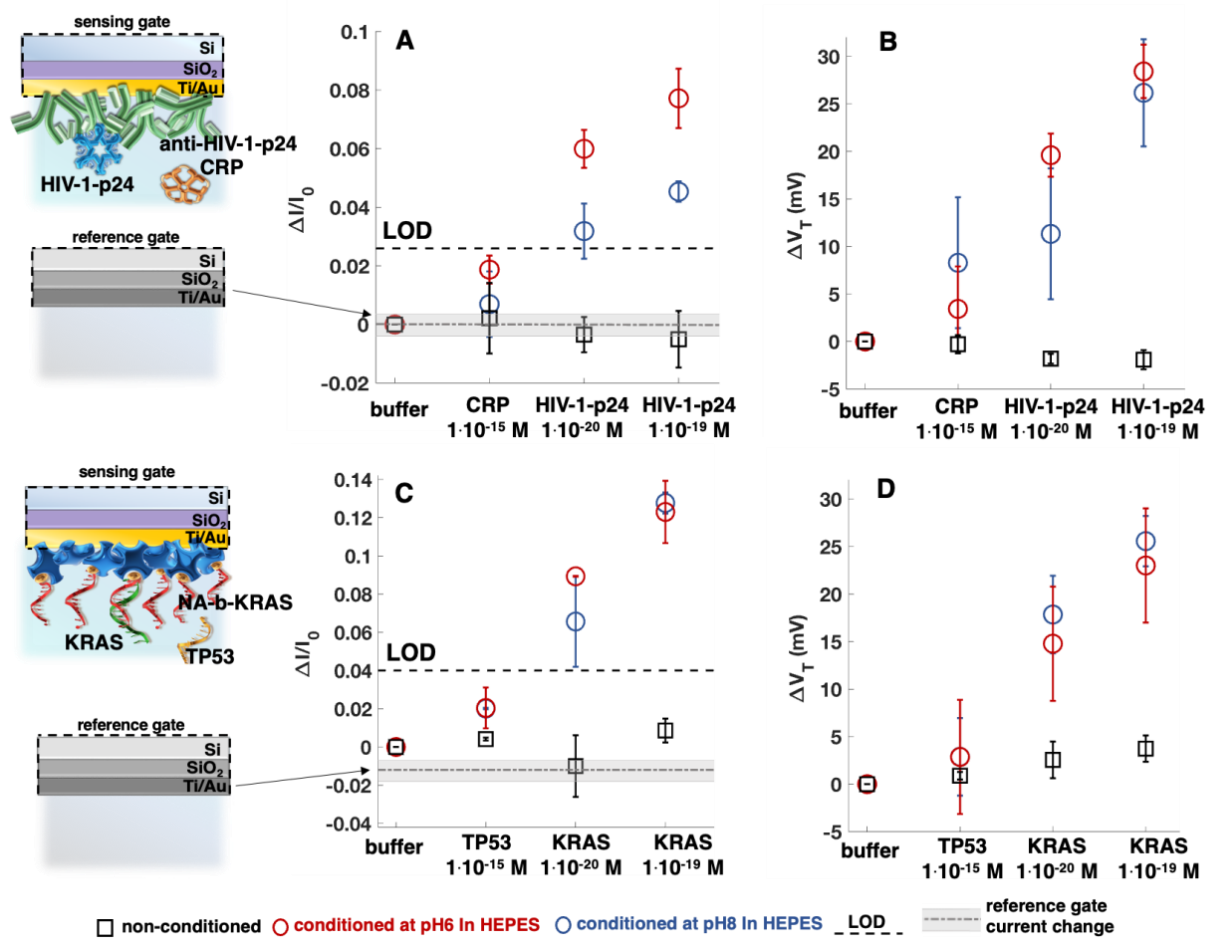

**Fig. 2**. **EGOFET source-drain current ($I_D$) and threshold voltage ($V_T$) shifts upon single/few-molecules sensing of HIV-1-p24 and KRAS** - $I_D$ ($V_{GS}$ = -0.5V and $V_D$ = -0.4V) current relative shifts $\Delta I/I_0$ [($I_D$-$I_0$)/$I_0$, $I_0$ being the baseline measured after incubation in the bare HEPES@pH7.4] averaged over two replicates, are measured in HEPES@pH7.4/$i_s$-low ($i_s$ = 5 mM) after incubation of the sensing gate in the higher salinity HEPES@pH7.4 ($i_s$ = 150 mM) solutions to be assayed. The $\Delta I/I_0$ and $\Delta V_T$ responses of the non-conditioned biolayers (black squares), and of the biolayers conditioned at pH 6 (red circles) or pH 8 (blue circles), are measured with the sensing gate at each sensing step. The dashed-dotted gray lines

represent the average $\Delta I/I_0$ measured with the reference electrodes, while the shaded areas represent one standard deviation. **(A)** and **(B)**: $\Delta I/I_0$ and $\Delta V_T$ for the HIV-1-p24 single/few-molecules sensing at the anti-HIV-1-p24 gate, with CRP serving as negative control (raw data in **figs. S24-S26**). **(C)** and **(D)**: $\Delta I/I_0$ and $\Delta V_T$ for the KRAS single/few-molecules sensing at the NA-b-KRAS gate, with TP53 serving as negative control (raw data in **figs. S27-S29**).

A literature search (**tab. S4**) shows that several large-area FETs capable of detecting below $10^{-15}$-$10^{-12}$ M, (or equivalently, involving a ratio between target-analytes and recognition-elements numerosity, of at least $10^4$), encompass a pH conditioning of the detecting electronic surface. This occurs during the necessary transition from the incubating solutions (at physiological pH and salinity) to the measuring electrolyte solutions with lower salinity and/or pH, in the intention of maximizing $\lambda_D$ and $I_D$. In this study the gate of the non-conditioned EGOFET devices is intentionally not subjected to any pH change. The responses falling below the LODs clearly demonstrate how this pH-conditioning solely enables the ultra-low LODs and not the often-evoked amplification associated with a FET-transduction. The latter, which affects both the noise (negative control) and the sensing signal, has little to no effect on lowering the LOD (*37*).

KPFM (*23*, *39*) images also quantify the $\Phi_S$ and $\sigma_S$ of the inspected antibodies and protein-probe complexes biolayers. The apparatus is schematically shown in **fig. S30** along with an energy diagram for the $\Delta\Phi_S$ shift upon single-molecule binding. This is equal to the corresponding $\Delta V_T$ shift, when measured within the same $\lambda_D$, scaling proportionally otherwise.

In **fig. S31** the Atomic Force Microscopy (AFM) topographical images illustrate a 1 $\mu m^2$ area of the three as deposited capturing and protein-probe biolayers at the sharp interface with an exposed portion of the substrate. The layers exhibit a densely packed coverage with contiguous structures much larger than an individual capturing or protein-probe element. This suggests the presence of interconnected aggregates of antibodies or protein-probe complexes that form already during physisorption (*40*, *41*).

In **Fig. 3A-3N** the KPFM images featuring the surface potential difference (SPD) across the interface between the biolayer and the substrate (serving as $\Phi_S$ reference level), are shown along with the relevant histograms. Here, the anti-HIV-1-p24 capturing layers, non-conditioned or conditioned at pH 6 or pH 8, are inspected. These images are captured on a large 90 x 90 $\mu m^2$ area, which accommodates approximately $10^8$ recognition elements. After incubation, at each step of the single/few-molecules sensing, the samples are washed with the low salinity HEPES@pH7.4/$i_s$-low buffer and dried. The same data are given for the anti-IgG on SiO$_2$ and NA-b-KARS on Au in **fig. S32** and **fig. S33**, respectively. In **Fig. 3O-3Q** the SPD shifts ($\Delta$SPD) values with respect to the corresponding negative control experiment, are summarized for all three biosystems.

The image in **Fig. 3B** shows a $(268 \pm 12)$ mV reduction of $\Phi_S$ on the sample portion covered with the HIV-1-p24 biolayer, with respect to the Au one, serving as reference. This is in qualitative agreement with the threshold voltage shift measured with the EGOFET and both are consistent with a solid surface zeta potential (*42*) assessment (**fig. S34**). The latter results in an isoelectric point of pH 3.6, indicating that at pH 7.4, the capturing layer is negatively charged. The $\Phi_S$ shift on the Na-b-KRAS layer is also negative, and so is its $\sigma_S$. Like the SPR and EGOFET data, the KPFM responses in the single/few-molecules regimes are beyond the LOD only on the pH-conditioned capturing/probe layers, with the pH 6 conditioned ones providing stronger responses. For the HIV-1-p24 sensing at $10^{-19}$ M (**Fig. 3M**) $\Delta$SPD is $106 \pm 16$ mV resulting in a relative increase (as compared to the baseline $\Phi_S$) of $(40 \pm 6)$ %, which is in quantitative agreement with the EGOFET data. Both EGOFET and KPFM potentiometric analyses show how after the sensing $\sigma_S$ is reduced while $\Phi_S$ increases in all the inspected biosystems.

The KPFM data provide an additional crucial insight: the images clearly demonstrate that, only on the pH-conditioned samples, one or few binding events trigger the change in the surface potential of a large sample area hosting $10^8$ recognition elements. This means that upon conditioning, each binding event can shift the surface potential of hundred-millions of other capturing antibodies or protein-probe complexes. Thus, widely propagating the effects of the conformational change that affects only the antibody or protein-probe complex directly involved in the affinity binding. Moreover, the extended surface potential shift is also independent of the presence of a metallic substrate.

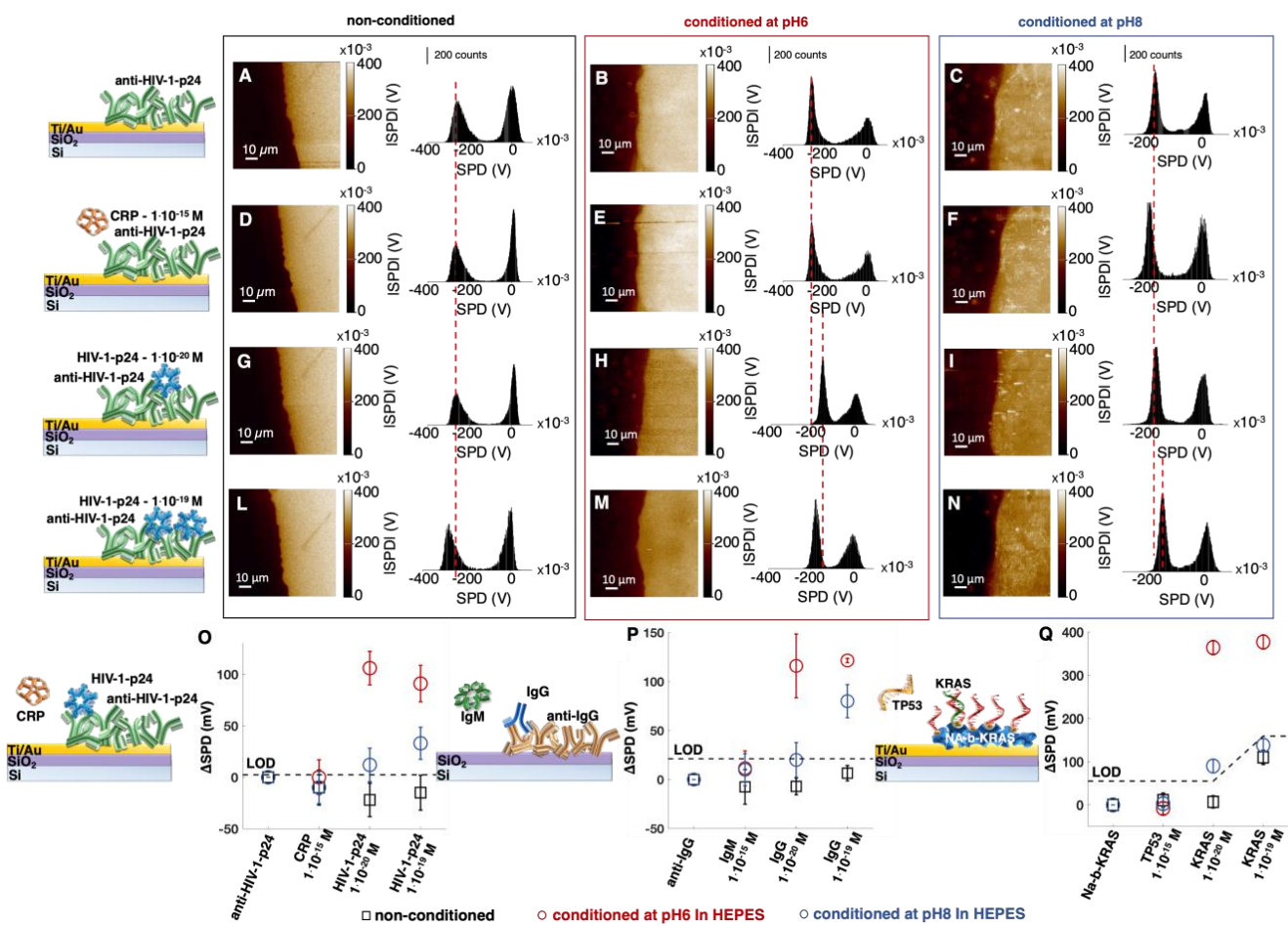

**Fig. 3. KPFM images of surface potential difference shifts upon single/few molecules sensing of two antigens and a DNA strand -** The surface potential differences (SPDs) of anti-HIV-1-p24 capturing layers at different steps of the HIV-1-p24 sensing protocol are imaged for non-conditioned capturing biolayers ({**A,D,G,L**} set of panels, black frame) or on the biolayers conditioned at pH 6 ({**B,E,H,M**}, red frame), and pH 8 ({**C,F,I,N**}, blue frame). Images 90 x 90 μm² are taken at each sensing step (first-row: baseline in HEPES@pH7.4; second-row: negative control experiment; third-row and forth-row: single-molecule and few-molecules sensing). The measurements are carried out in air after washing in HEPES@pH7.4/$i_s$-low. The samples are patterned to create a sharp interface between the biolayer and the substrate, serving as an internal standard. Each panel comprises both the image and the relevant histogram distribution. ΔSPD (SPD shifts with respect to the baseline) average values are plotted at the different

stages of the sensing protocol for the anti-HIV-1-p24 (**O**), the anti-IgG (raw data **fig. S32**) (**P**), and the NA-b-KRAS (raw data **fig. S33**) (**Q**) biosystems, on non-conditioned (black squares), conditioned at pH 6 (red circles) or pH 8 (blue circles).

**A phenomenological sensing mechanism**

To unravel the mechanism of the single/few-molecules sensing at a large detecting interface, additional pieces of evidence are collected. The degree of hydrophilicity of a 1 cm$^2$ anti-IgG capturing layer surface is measured with a Static Contact Angle (SCA) system (**fig. S35**) upon conditioning and at each sensing step (**fig. S36**, **Fig. 4A**). Once more the key role of the pH-conditioning in enabling sensing at the physical limit, is proven with an SCA reaching 61 ± 4 (°) in the few-molecule sensing regime on an anti-IgG biolayer conditioned at pH 6. This is an 18 ± 7 (°) increase as compared to a negative control SCA of 43 ± 7 (°). No appreciably reliable signal is recorded on upon conditioning at pH 8.

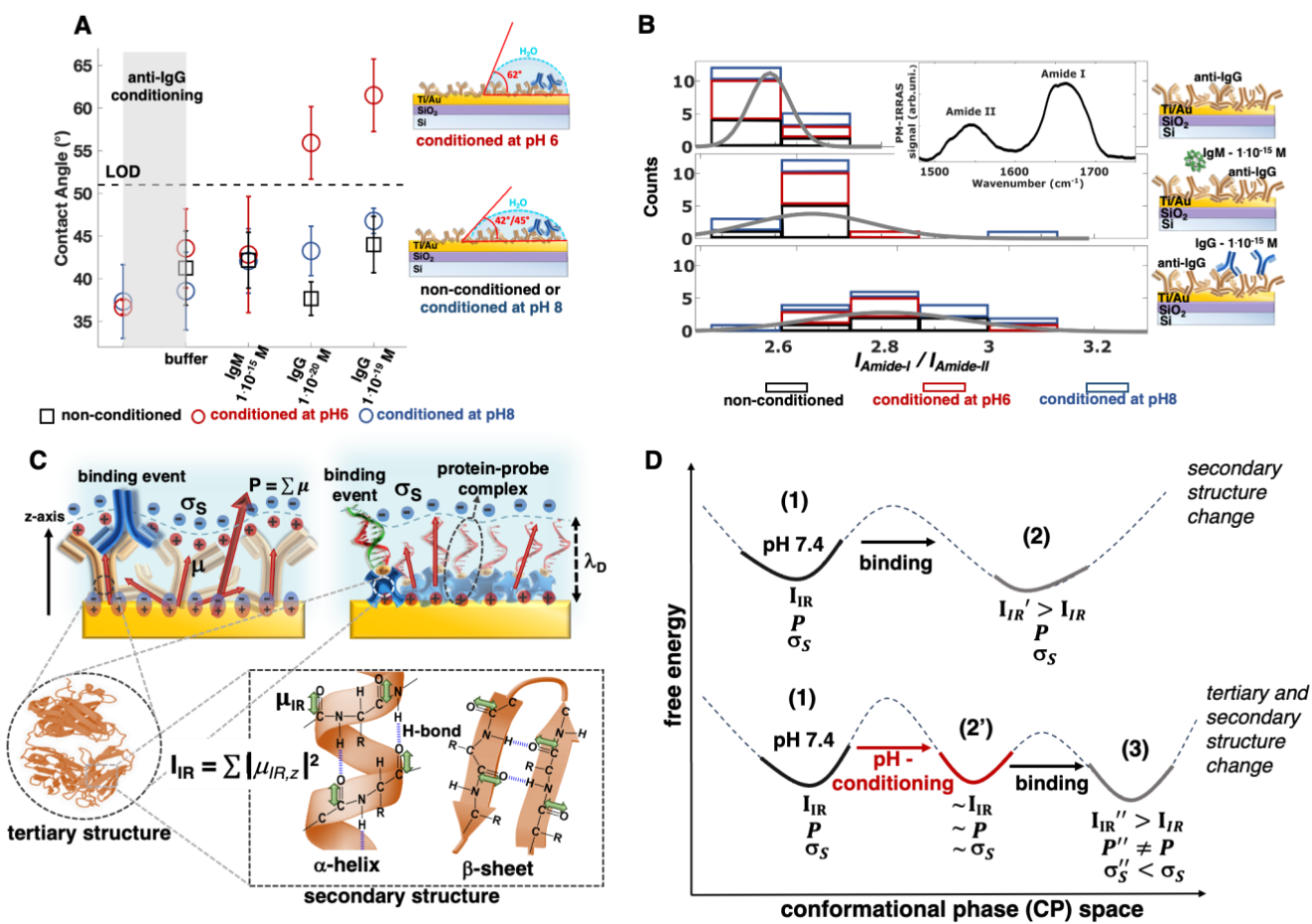

**Fig. 4. Sensing mechanism -** (**A**) Average static contact angles (SCA, three replicates) measured at each sensing step on non-conditioned (black squares), conditioned at pH 6 (red circles), or at pH 8 (blue circles) anti-IgG biolayers. The first data points (from the left) are relevant to the pristine as deposited anti-IgG layers that undergo a pH-conditioning or not. Afterward, the samples to the standard sensing steps. (**B**) Distribution of the Amide I and Amide II peak integrated areas ratios, $I_{Amide-I} / I_{Amide-II}$, measured on 54 samples (**tab. S5**). A typical PM-IRRAS spectrum in the 1480 - 1750 cm$^{-1}$ amide region is shown in the inset. In the top row, baseline data taken from the anti-IgG layers are displayed, in the middle row

the spectra are taken from the anti-IgG exposed to IgM $1 \cdot 10^{-15}$ M (negative control experiment), while the bottom row represents the IgG sensing at $1 \cdot 10^{-15}$ M. Black, red, and blue boxes in the histograms indicate the samples conditioned according to the color code. The gray line is the gaussian fit of the data. (**C**) Schematic representation of the structure of a capturing layer sensing its antigen (top-left) and of a protein-probe complex sensing its complementary DNA target (top-right). The insets at the bottom feature the tertiary and the secondary molecular structure of a generic protein. (**D**) Energy landscapes of conformational phases (CP) in the phenomenological model for the mechanism of the single/few-molecules at a large interface.

The increase in the hydrophobic character is complemented by a shift in the nanomechanical properties of an anti-IgG layer conditioned at a pH of ~ 5.5 (**fig. S37**). Upon binding of few antigens (antigens/capturing-antibodies ratio of $10^6$), a shift in both Young's modulus, decreasing by $(37 \pm 8)$ %, and of the adhesion force distributions, increasing by $(66 \pm 12)$ %, are measured.

Another piece of relevant information is provided by a PM-IRRAS (*43*) investigation. A schematic view of the apparatus is in **fig. S38**. In **Fig. 4B** the distributions of peak area ratios $I_{Amide-I}/I_{Amide-II}$ at the different sensing steps, are shown. This ratio, assessed on bovine serum albumin, reflects changes in a protein secondary structure, and it quantitatively increases with its α-helix content (*44*). Our data show that $I_{Amide-I}/I_{Amide-II}$ is $2.59 \pm 0.04$ for the baseline samples, it slightly increases in the negative control experiments ($2.67 \pm 0.11$) and reaches the value $2.81 \pm 0.15$ upon affinity binding. In agreement with the literature (*45*), no sizable dependance on the pH-conditioning is seen. Also in these experiments, the changes are triggered by a density of affinity binding as low as ~ 1 in $10^8$ available capturing antibodies. The reliability of these results is tested with an artificial-intelligence-based data analysis (**fig. S39** and **fig. S40**).

A biological recognition layer is schematically depicted in **Fig. 4C** featuring a monolayer of capturing antibodies (on the left) and of protein-probe complexes (on the right) along with their dipole moments **μ** (*46*) contributing to $\mathbf{P} = \sum \boldsymbol{\mu}$ and to a negative $\sigma_s$. The tertiary and secondary structures of a generic protein are illustrated in the insets, evidencing the molecular backbones and the three-dimensional conformation structures, α-helix and β-sheet. Additionally, the transition dipole moments $\mu_{IR}$ whose squared z-component additively contribute to the PM-IRRAS signal $I_{IR} = \sum |\mu_{IR,z}|^2$, are also displayed.

Proteins dipole-dipole and dipole-substrate interactions lead to a not totally disordered physisorbed layer (*47*) of partially unfolded proteins (*48*). When proteins partially unfold, they lose their native structure exposing hydrophobic functionalities, which makes them more prone to interact forming aggregates (*24*). Such a state is identified as the relative minimum (**1**) in the conformational phase (CP) space of **Fig. 4D**. Few affinity bindings stabilize the biolayer populated by trillions of recognition elements in CP (**2**). Here, the increase of the $I_{Amide-I}/I_{Amide-II}$ ratio, associated to no **P** and $\sigma_s$ appreciable changes, indicates that the affinity binding affects the secondary structure. This also indicates a growth of secondary structures with an aggregated three-dimensional conformation, possibly at the intramolecular level. Conditioning at an acidic pH (*49*) involves partially unfolded proteins in aggregation pathways to a further extent, leading also to the formation of transient species such as for instance the molted globules (*50*, *51*). The likely metastable CP (**2'**) achieved by conditioning at pH 6 (and to a lesser extent at pH 8), is characterized by small measurable changes. However, upon single/few-molecules affinity binding the layer converts to CP (**3**), where the surface becomes visibly more hydrophobic. The more pronounced hydrophobic character makes them also stickier (*52*, *53*), more flexible and less stiff (*54*) as the Young's modulus decreases (*49*). The measured increase of the hydrophobization, results also in an expected and reliably measurable reduction in $\sigma_s$. The corresponding $\Phi_S$ increase upon single-molecule binding, is also proven to spread over a large surface hosting hundreds of millions of recognition elements. Importantly, the single/few-molecules sensing here proposed induces only a partial denaturation, as the pH-conditioned layers can still perform affinity binding at $10^{-7}$ M concentrations (double-layer regime).

This work demonstrates how single/few-molecules affinity bindings on a pH-conditioned recognition layer, formed by capturing antibodies or protein-probe complexes, lead to a collective rearrangement of its secondary and tertiary structures. The process is initiated by the formation of a single or a few affinity couples, and their conformational changes propagate at least to hundreds of millions of other antibodies or protein-probe complexes populating the detecting surface. This amplification process shifts the surface potential of an area many orders of magnitudes larger than the footprint of the triggering target, enabling reliable single-molecule detection of both proteins and nucleic-acids, at $10^{-20}$ M.

Such findings can pave the way for improved control over the formation of protein aggregates, deepening our understanding and enhancing the prospects for curing proteins-misfolding related diseases (*55*). The impact is also anticipated in the rapidly expanding field of portable devices, both plasmonic and electronic, which offer improved accountability in single biomarker-based diagnostics with high accuracy for point-of-care settings (*13*).

**References and Notes**

1. C. Alix-Panabieres, Perspective: The future of liquid biopsy. *Nature* **579**, S9 (2020).
2. J. D. Cohen, L. Li, Y. Wang, C. Thoburn, B. Afsari, L. Danilova, C. Douville, A. A. Javed, F. Wong, A. Mattox, R. H. Hruban, C. L. Wolfgang, M. G. Goggins, M. Dal Molin, T.-L. Wang, R. Roden, A. P. Klein, J. Ptak, L. Dobbyn, J. Schaefer, N. Silliman, M. Popoli, J. T. Vogelstein, J. D. Browne, R. E. Schoen, R. E. Brand, H.-L. Wong, A. S. Mansfield, J. Jen, S. M. Hanash, M. Falconi, P. J. Allen, S. Zhou, C. Bettegowda, L. A. Diaz Jr, C. Tomasetti, K. W. Kinzler, B. Vogelstein, A. Marie Lennon, N. Papadopoulos, Detection and localization of surgically resectable cancers with a multi-analyte blood test. *Science* **359**, 926–930 (2018).
3. A. Erlich, D. Gelfand, J. J. Sninsky, Recent Advances in the Polymerase Chain Reaction. *Science* **252**, 1643 (1991).
4. J. M. Rothberg, W. Hinz, T. M. Rearick, J. Schultz, W. Mileski, M. Davey, J. H. Leamon, K. Johnson, M. J. Milgrew, M. Edwards, J. Hoon, J. F. Simons, D. Marran, J. W. Myers, J. F. Davidson, A. Branting, J. R. Nobile, B. P. Puc, D. Light, T. A. Clark, M. Huber, J. T. Branciforte, I. B. Stoner, S. E. Cawley, M. Lyons, Y. Fu, N. Homer, M. Sedova, X. Miao, B. Reed, J. Sabina, E. Feierstein, M. Schorn, M. Alanjary, E. Dimalanta, D. Dressman, R. Kasinskas, T. Sokolsky, J. A. Fidanza, E. Namsaraev, K. J. McKernan, A. Williams, G. T. Roth, J. Bustillo, An integrated semiconductor device enabling non-optical genome sequencing. *Nature* **475**, 348–352 (2011).
5. L. Cohen, D. R. Walt, Single-molecule arrays for protein and nucleic acid analysis. *Annual Review of Analytical Chemistry* **10**, 345–363 (2017).
6. J. Kaiser, 'Liquid biopsy' for cancer promises early detection: Combining DNA and protein markers brings researchers closer to a universal cancer screening test. *Science* **359**, 259 (2018).
7. B. D. Reed, M. J. Meyer, V. Abramzon, O. Ad, P. Adcock, F. R. Ahmad, G. Alppay, J. A. Ball, J. Beach, D. Belhachemi, A. Bellofiore, M. Bellos, J. F. Beltrán, A. Betts, M. Wadud Bhuiya, K. Blacklock, R. Boer, D. Boisvert, N. D. Brault, A. Buxbaum, S. Caprio, C. Choi, T. D. Christian, R. Clancy, J. Clark, T. Connolly, K. Fink Croce, R. Cullen, M. Davey, J. Davidson, M. M. Elshenawy, M. Ferrigno, D. Frier, S. Gudipati, S. Hamill, Z. He, S. Hosali, H. Huang, L. Huang, A. Kabiri, G. Kriger, B. Lathrop, A. Li, P. Lim, S. Liu, F. Luo, C. Lv, X. Ma, E. Mccormack, M. Millham, R. Nani, M. Pandey, J. Parillo, G. Patel, D. H. Pike, K. Preston, A. Pichard-Kostuch, K. Rearick, T. Rearick, M. Ribezzi-Crivellari, G. Schmid, J. Schultz, X. Shi, B. Singh, N. Srivastava, S. F. Stewman, T. R. Thurston, P. Trioli, J. Tullman, X. Wang, Y.-C. Wang, E. A. G. Webster, Z. Zhang, J. Zuniga, S. S. Patel, A. D. Griffiths, A. M. Van Oijen, M. Mckenna, M. D. Dyer, J. M. Rothberg, Real-time dynamic single-molecule protein sequencing on an integrated semiconductor device. *Science* **378**, 186–192 (2022).
8. J. A. Alfaro, P. Bohländer, M. Dai, M. Filius, C. J. Howard, X. F. van Kooten, S. Ohayon, A. Pomorski, S. Schmid, A. Aksimentiev, E. V. Anslyn, G. Bedran, C. Cao, M. Chinappi, E. Coyaud, C. Dekker, G. Dittmar, N. Drachman, R. Eelkema, D. Goodlett, S. Hentz, U. Kalathiya, N. L. Kelleher, R. T. Kelly, Z. Kelman, S. H. Kim, B. Kuster, D. Rodriguez-Larrea, S. Lindsay, G. Maglia, E. M. Marcotte, J. P. Marino, C. Masselon, M. Mayer, P. Samaras, K. Sarthak, L. Sepiashvili, D. Stein, M. Wanunu, M. Wilhelm, P. Yin, A. Meller, C. Joo, The emerging landscape of single-molecule protein sequencing technologies. *Nat Methods* **18**, 604–617 (2021).

9. H. Brinkerhoff, A. S. W. Kang, J. Liu, A. Aksimentiev, C. Dekker, Multiple rereads of single proteins at single-amino acid resolution using nanopores. *Science* **374**, 1509–1513 (2021).
10. B. M. Floyd, E. M. Marcotte, Protein Sequencing, One Molecule at a Time. *Annu. Rev. Biophys.* **51**, 181–200 (2022).
11. D. M. Rissin, C. W. Kan, T. G. Campbell, S. C. Howes, D. R. Fournier, L. Song, T. Piech, P. P. Patel, L. Chang, A. J. Rivnak, E. P. Ferrell, J. D. Randall, G. K. Provuncher, D. R. Walt, D. C. Duffy, Single-molecule enzyme-linked immunosorbent assay detects serum proteins at subfemtomolar concentrations. *Nat Biotechnol* **28**, 595–599 (2010).
12. E. Macchia, K. Manoli, B. Holzer, C. Di Franco, M. Ghittorelli, F. Torricelli, D. Alberga, G. F. Mangiatordi, G. Palazzo, G. Scamarcio, L. Torsi, Single-molecule detection with a millimetre-sized transistor. *Nat Commun* **9** (2018), https://doi.org/10.1038/s41467-018-05235-z.
13. E. Macchia, F. Torricelli, M. Caputo, L. Sarcina, C. Scandurra, P. Bollella, M. Catacchio, M. Piscitelli, C. Di Franco, G. Scamarcio, L. Torsi, Point-Of-Care Ultra-Portable Single-Molecule Bioassays for One-Health. *Advanced Materials*, doi: 10.1002/adma.202309705 (2023).
14. E. Genco, F. Modena, L. Sarcina, K. Björkström, C. Brunetti, M. Caironi, M. Caputo, V. M. Demartis, C. Di Franco, G. Frusconi, L. Haeberle, P. Larizza, M. T. Mancini, R. Österbacka, W. Reeves, G. Scamarcio, C. Scandurra, M. Wheeler, E. Cantatore, I. Esposito, E. Macchia, F. Torricelli, F. A. Viola, L. Torsi, A Single-Molecule Bioelectronic Portable Array for Early Diagnosis of Pancreatic Cancer Precursors. *Advanced Materials* **35** (2023).
15. E. Macchia, Z. M. Kovács-Vajna, D. Loconsole, L. Sarcina, M. Redolfi, M. Chironna, F. Torricelli, L. Torsi, A handheld intelligent single-molecule binary bioelectronic system for fast and reliable immunometric point-of-care testing. *Sci Adv* **8**, 1–15 (2022).
16. Y. Wu, D. Bennett, R. D. Tilley, J. J. Gooding, How Nanoparticles Transform Single Molecule Measurements into Quantitative Sensors. *Advanced Materials* **32** (2020).
17. J. Sotres, H. Boyd, J. F. Gonzalez-Martinez, Enabling autonomous scanning probe microscopy imaging of single molecules with deep learning. *Nanoscale* **13**, 9193–9203 (2021).
18. R. Ren, Y. Zhang, B. P. Nadappuram, B. Akpinar, D. Klenerman, A. P. Ivanov, J. B. Edel, Y. Korchev, Nanopore extended field-effect transistor for selective single-molecule biosensing. *Nat Commun* **8** (2017).
19. R. Chikkaraddy, B. De Nijs, F. Benz, S. J. Barrow, O. A. Scherman, E. Rosta, A. Demetriadou, P. Fox, O. Hess, J. J. Baumberg, Single-molecule strong coupling at room temperature in plasmonic nanocavities. *Nature* **535**, 127–130 (2016).
20. M. Thompson, S. L. R. Ellison, R. Wood, Harmonized guidelines for single-laboratory validation of methods of analysis (IUPAC Technical Report). *Pure and Applied Chemistry* **74**, 835–855 (2002).
21. W. Knoll, Interfaces and thin films as seen by bound electromagnetic waves. *Annu Rev Phys Chem* **49**, 569–638 (1998).
22. F. Torricelli, D. Z. Adrahtas, F. Biscarini, A. Bonfiglio, C. A. Bortolotti, C. D. Frisbie, I. Mcculloch, E. Macchia, G. G. Malliaras, Electrolyte- gated transistors for enhanced performance bioelectronics. *Nature Reviews Methods Primers* **1** (2021), https://doi.org/10.1038/s43586-021-00071-w.
23. M. Nonnenmacher, M. P. O'Boyle, H. K. Wickramasinghe, Kelvin probe force microscopy. *Appl Phys Lett* **58**, 2921–2923 (1991).

24. W. Wang, S. Nema, D. Teagarden, Protein aggregation-Pathways and influencing factors. *Int J Pharm* **390**, 89–99 (2010).
25. E. E. Meyer, K. J. Rosenberg, J. Israelachvili, Recent progress in understanding hydrophobic interactions. *PNAS* **103**, 15739–15746 (2006).
26. H. H. Nguyen, J. Park, S. Kang, M. Kim, Surface plasmon resonance: A versatile technique for biosensor applications. *Sensors (Switzerland)* **15**, 10481–10510 (2015).
27. T. Neumann, M. L. Johansson, D. Kambhampati, W. Knoll, Surface-plasmon fluorescence spectroscopy. *Adv Funct Mater* **12**, 575–586 (2002).
28. J. N. Anker, W. P. Hall, O. Lyandres, N. C. Shah, J. Zhao, R. P. Van Duyne, Biosensing with plasmonic nanosensors. *Nanoscience and Technology: A Collection of Reviews from Nature Journals* **7**, 308–319 (2009).
29. S. Y. Ding, J. Yi, J. F. Li, B. Ren, D. Y. Wu, R. Panneerselvam, Z. Q. Tian, Nanostructure-based plasmon-enhanced Raman spectroscopy for surface analysis of materials. *Nat Rev Mater* **1** (2016), https://doi.org/10.1038/natrevmats.2016.21.
30. T. Špringer, Z. Krejčík, J. Homola, Detecting attomolar concentrations of microRNA related to myelodysplastic syndromes in blood plasma using a novel sandwich assay with nanoparticle release. *Biosens Bioelectron* **194** (2021), https://doi.org/10.1016/j.bios.2021.113613.
31. L. Sarcina, C. Scandurra, C. Di Franco, M. Caputo, M. Catacchio, P. Bollella, G. Scamarcio, E. Macchia, L. Torsi, A stable physisorbed layer of packed capture antibodies for high-performance sensing applications. *J Mater Chem C Mater* **11** (2023), https://doi.org/10.1039/d3tc01123b.
32. M. Pichlo, S. Bungert-Plümke, I. Weyand, R. Seifert, W. Bönigk, T. Strünker, N. D. Kashikar, N. Goodwin, A. Müller, P. Pelzer, Q. Van, J. Enderlein, C. Klemm, E. Krause, C. Trötschel, A. Poetsch, E. Kremmer, U. B. Kaupp, High density and ligand affinity confer ultrasensitive signal detection by a guanylyl cyclase chemoreceptor. *Journal of Cell Biology* **206**, 541–557 (2014).
33. T. L. McMeekin, M. L. Groves, N. J. Hipp, Refractive Indices of Amino Acids, Proteins, and Related Substances. *In Amino Acids and Serum Proteins; Stekol, J.; Advances in Chemistry; American Chemical Society: Washington, DC*, 54–66 (1964).
34. D. Khago, J. C. Bierma, K. W. Roskamp, N. Kozlyuk, R. W. Martin, Protein refractive index increment is determined by conformation as well as composition. *Journal of Physics Condensed Matter* **30** (2018), https://doi.org/10.1088/1361-648X/aae000.
35. H. Šípová-Jungová, L. Jurgová, K. Mrkvová, N. S. Lynn, B. Špačková, J. Homola, Biomolecular charges influence the response of surface plasmon resonance biosensors through electronic and ionic mechanisms. *Biosens Bioelectron* **126**, 365–372 (2019).
36. M. Kaisti, Detection principles of biological and chemical FET sensors. *Biosens Bioelectron* **98**, 437–448 (2017).
37. E. Macchia, F. Torricelli, P. Bollella, L. Sarcina, A. Tricase, C. Di Franco, R. Österbacka, Z. M. Kovács-Vajna, G. Scamarcio, L. Torsi, Large-Area Interfaces for Single-Molecule Label-free Bioelectronic Detection. *Chem Rev* **122**, 4636–4699 (2022).
38. L. Kergoat, L. Herlogsson, B. Piro, M. C. Pham, G. Horowitz, X. Crispin, M. Berggren, Tuning the threshold voltage in electrolyte-gated organic field-effect transistors. *PNAS* **109,** 8394-8399 (2012).
39. C. Di Franco, E. Macchia, L. Sarcina, N. Ditaranto, A. Khaliq, L. Torsi, G. Scamarcio, Extended Work Function Shift of Large-Area Biofunctionalized Surfaces Triggered by a Few Single-Molecule Affinity Binding Events. *Adv Mater Interfaces* **6** (2022), https://doi.org/10.1002/admi.202201829.

40. G. P. Gorbenko, P. K. J. Kinnunen, The role of lipid-protein interactions in amyloid-type protein fibril formation. *Chem Phys Lipids* **141**, 72–82 (2006).
41. F. Liu, X. Li, A. Sheng, J. Shang, Z. Wang, J. Liu, Kinetics and Mechanisms of Protein Adsorption and Conformational Change on Hematite Particles. *Environ Sci Technol* **53**, 10157–10165 (2019).
42. H. Buksek, T. Luxbacher, I. Petrinic, Zeta Potential Determination of Polymeric Materials Using Two Differently Designed Measuring Cells of an Electrokinetic Analyzer. *Acta Chim. Slov* **57**, 700–706 (2010).
43. B. L. Frey, R. M. Corn, S. C. Weibel, "Polarization-Modulation Approaches to Reflection–Absorption Spectroscopy" in *Handbook of Vibrational Spectroscopy* (Wiley, 2001).
44. Kato K. Matsui T. Tanaka S., Quantitative Estimation of a-Helix Coil Content in Bovine Serum Albumin by Fourier Transform-Infrared Spectroscopy. *Appl Spectrosc* **41**, 861–864 (1987).
45. Kenneth P. Ishida and Peter R. Griffiths, Comparison of the Amide I/II Intensity Ratio of Solution and Solid-State Proteins Sampled by Transmission, Attenuated Total Reflectance, and Diffuse Reflectance Spectrometry. *Appl Spectrosc* **47**, 584–589 (1993).
46. S. N. Singh, S. Yadav, S. J. Shire, D. S. Kalonia, Dipole-dipole interaction in antibody solutions: Correlation with viscosity behavior at high concentration. *Pharm Res* **31**, 2549–2558 (2014).
47. S. Emaminejad, M. Javanmard, C. Gupta, S. Chang, R. W. Davis, R. T. Howe, Tunable control of antibody immobilization using electric field. *Proc Natl Acad Sci U S A* **112**, 1995–1999 (2015).
48. R. A. Latour, Fundamental Principles of the Thermodynamics and Kinetics of Protein Adsorption to Material Surfaces. *Colloids Surf B Biointerfaces* **191** (2020), https://doi.org/10.1016/j.colsurfb.2020.110992.
49. M. L. Donten, P. Hamm, PH-jump induced α-helix folding of poly-L-glutamic acid. *Chem Phys* **422**, 124–130 (2013).
50. V. Filipe, B. Kükrer, A. Hawe, W. Jiskoot, Transient molten globules and metastable aggregates induced by brief exposure of a monoclonal IgG to Low pH. *J Pharm Sci* **101**, 2327–2339 (2012).
51. Redfield C Smith RAG Dobson CM, Structural characterization of a highly–ordered 'molten globule' at low pH. *Nat Struct Biol* **1**, 23–29 (1994).
52. J. H. M. Van Gils, D. Gogishvili, J. Van Eck, R. Bouwmeester, E. Van Dijk, S. Abeln, How sticky are our proteins? Quantifying hydrophobicity of the human proteome. *Bioinformatics Advances* **2**, 1–9 (2022).
53. S. N. Jamadagni, R. Godawat, S. Garde, Hydrophobicity of proteins and interfaces: Insights from density fluctuations. *Annu Rev Chem Biomol Eng* **2**, 147–171 (2011).
54. L. R. Khoury, I. Popa, Chemical unfolding of protein domains induces shape change in programmed protein hydrogels. *Nat Commun* **10** (2019), https://doi.org/10.1038/s41467-019-13312-0.
55. M. Jucker, L. C. Walker, Propagation and spread of pathogenic protein assemblies in neurodegenerative diseases. *Nat Neurosci* **21**, 1341–1349 (2018).